\documentclass[useAMS,usenatbib]{mn2e}
\newcommand \kms{km s$^{-1}$}
\newcommand \zabs{z$_{\rm abs}$}
\newcommand \zem{z$_{\rm em}$}
\newcommand \url{}

\usepackage{longtable}
\usepackage{graphicx}
\usepackage{lscape}
\usepackage{rotating}
\usepackage{amssymb}

\def \nodata{. . .}
\def \HI{H\textsc{i}}
\def \sion{\textsc{ii}}
\def \ndla{29}
\def \vninety{v$_{90}$}
\def \Mgeqw{W$_{0}^{2796}$}
\def \Feeqw{W$_{0}^{2600}$}
\def \fracMgFe{ ${\rm{W}_{0}^{2796}}$/${\rm{W}_{0}^{2600}}$}
\def \omegaDLA{$\Omega_{\rm H \textsc{i}}$}
\begin{document}

\title[Selection of DLAs with Mg\textsc{ii}]{On the selection of damped Lyman alpha systems using MgII absorption at \textbf{$2<$\zabs{}$<4$.}}
\author[Berg et al.] {
\parbox[t]{\textwidth}{
T. A. M. Berg$^1$,  S. L. Ellison$^1$, J. X. Prochaska$^2$, R. S\'anchez-Ram\'irez$^{3,4,5}$,  S. Lopez$^{6}$, V. D'Odorico$^{7}$, G. Becker$^{8,9}$, L. Christensen$^{10}$, G. Cupani$^{7}$, K. Denney$^{11}$, G. Worseck$^{12}$}\\\\
$^1$ Department of Physics and Astronomy, University of Victoria, Victoria, British Columbia, V8P 1A1, Canada.\\
$^2$ Department of Astronomy and Astrophysics, University of California, Santa Cruz, Santa Cruz, CA, 95064, USA.\\
$^3$Unidad Asociada Grupo Ciencias Planetarias (UPV/EHU, IAA-CSIC), Departamento de F\'isica Aplicada I,\\ E.T.S. Ingenier\'ia, Universidad del Pa\'is Vasco (UPV/EHU), Alameda de Urquijo s/n, E-48013 Bilbao, Spain.\\
$^4$Ikerbasque, Basque Foundation for Science, Alameda de Urquijo 36-5, E-48008 Bilbao, Spain.\\
$^5$Instituto de Astrof\'isica de Andaluc\'ia (IAA-CSIC), Glorieta de la Astronom\'ia s/n, E-18008, Granada, Spain.\\
$^{6}$Departamento de Astronom\'{\i}a, Universidad de Chile, Casilla 36-D, Santiago, Chile.\\
$^{7}$INAF-Osservatorio Astronomico di Trieste, Via Tiepolo 11, I-34143 Trieste, Italy.\\
$^{8}$Space Telescope Science Institute, 3700 San Martin Drive, Baltimore, MD 21218, USA.\\
$^{9}$Institute of Astronomy and Kavli Institute of Cosmology, Madingley Road, Cambridge CB3 0HA, UK.\\
$^{10}$Dark Cosmology Centre, Niels Bohr Institute, University of Copenhagen, Juliane Maries Vej 30, DK-2100 Copenhagen, Denmark.\\
$^{11}$Department of Astronomy, The Ohio State University, 140 West 18th Avenue, Columbus, OH 43210, USA.\\
$^{12}$Max-Planck-Institut f\"{u}r Astronomie, K\"{o}nigstuhl 17, D-69117 Heidelberg, Germany.\\
}

\maketitle

\begin{abstract}
The XQ-100 survey provides optical and near infrared coverage of 36 blindly selected, intervening damped Lyman $\alpha$ systems (DLAs) at $2<$\zabs{}$<4$, simultaneously covering the  Mg\sion{} doublet at $\lambda \lambda$ 2796, 2803 \AA{}, and the Ly$\alpha$ transition. Using the XQ-100 DLA sample, we investigate the completeness of selecting DLA absorbers based on their Mg\sion{} rest-frame equivalent width (\Mgeqw{}) at these redshifts.  Of the 29 DLAs with clean Mg\sion{} profiles, we find that six (20\% of DLAs) have \Mgeqw{}$<0.6$ \AA{}. The DLA incidence rate of \Mgeqw{}$<0.6$ \AA{} absorbers is a factor of $\sim5$ higher than what is seen in $z\sim1$ samples, indicating a potential evolution in the Mg\sion{} properties of DLAs with redshift. All of the \Mgeqw{}$<0.6$ \AA{} DLAs have low metallicities ($-2.5<[M/H]<-1.7$), small velocity widths (\vninety$<50$ \kms{}), and tend to have relatively low N(\HI{}). We demonstrate that the exclusion of these low \Mgeqw{} DLAs results in a higher mean N(\HI{}) which in turn leads to a $\sim7$\% increase in the cosmological gas density of \HI{} of DLAs at $2<$\zabs{}$<4$; and that this exclusion has a minimal effect on the \HI{}-weighted mean metallicity.
\end{abstract}


\section{Introduction}

Quasar (QSO) absorption line systems provide an excellent probe of the evolution of the \HI{} gas content over cosmic time. Of the many classes of QSO absorption line systems, damped Lyman $\alpha$ systems (DLAs) are the highest \HI{} column density absorbers, defined as having log N(\HI{})$\geq 20.3$ \citep{Wolfe86,Wolfe05}. Although fewer in number compared to lower N(\HI{}) counterparts (such as subDLAs; $19.0<$logN(\HI{})$<20.3$), DLAs dominate the \HI{} column density distribution from \zabs{}$\sim5$ to the present epoch  and are used to trace the cosmological gas density of \HI{} (\omegaDLA{}), eventually fuelling future generations of star formation \citep[][]{Lanzetta95,Rao00,StorrieLombardi00,Peroux03,Prochaska05DR3,Rao06,Prochaska09,
Noterdaeme12,Zafar13,Crighton15, Neeleman16,SanchezRamirez16}. At absorption redshifts where the \HI{} is observed in optical bands (\zabs{}$\gtrsim1.5$), \omegaDLA{} remains relatively constant with redshift \citep[for the most recent results at these redshifts, see][]{Crighton15,SanchezRamirez16}.  At $z\sim 0$, \omegaDLA{} is currently best measured from 21 cm emission line surveys of galaxies \citep{Zwaan05,Martin10}. Between these $z\sim 0$ measurements and \omegaDLA{} measured in DLAs at $z \sim 1.5$, the gas content of galaxies has only evolved by a  factor of $\sim2$ \citep{Zwaan05, SanchezRamirez16}.

Despite well constrained estimates of \omegaDLA\ at $z \sim 0$ and at $z>2$, studying the nature of the \omegaDLA{} evolution between $0.3\lesssim$\zabs{}$\lesssim1.5$ is challenging, as the Ly$\alpha$ transition shifts into the ultraviolet, requiring expensive space-based observations; and 21 cm emission becomes extremely difficult to detect \citep{Rhee16}.  In an effort to improve the efficiency of space telescope observations, it has become common practice to pre-select candidate DLAs based on the rest-frame equivalent widths (EWs) of the associated Mg\sion{} $\lambda \lambda$ 2796, 2803 \AA{} absorption observed in the optical \citep[][hereafter referred to as R00 and R06, respectively]{Rao00,Rao06}. With the inclusion of absorbers satisfying a Mg\sion{} 2796 \AA{} EW cut of \Mgeqw{}$\geq0.3$ \AA{} (R00), the final statistical sample compiled in R06 contains \emph{no} DLAs at \zabs{}$\sim1$ with \Mgeqw{}$<0.6$ \AA{}\footnote{ Although some DLAs with \Mgeqw{}$<0.6$ \AA{} have been previously identified \citep[e.g.][R06]{Peroux04}.}.

\omegaDLA{} derived from \zabs{}$\sim1$  DLA samples pre-selected from Mg\sion{} (\omegaDLA{}$\sim7.5\times10^{-3}$) are consistent with the \zabs{}$\gtrsim 2$ value, implying strong evolution at the lowest redshifts (R06). However, a recent `blind' archival survey of DLAs at $z\sim1$ derived a value of \omegaDLA{} a factor of 3 lower than R06 ($\sim2.5\times10^{-3}$), and consistent with 21 cm results at $z\sim0$ \citep{Neeleman16}. This tension in \omegaDLA{} has led to suggestions that Mg\sion{} DLA pre-selection may be biased, possibly leading to high \omegaDLA{} \citep[][]{Peroux04,Zavadsky09,Neeleman16}.

In this Letter, we investigate the nature of Mg\sion{} selection of 36 DLAs at $2<$\zabs{}$<4$ from the XQ-100 Legacy Survey (P.I. S. Lopez). The blind nature of the XQ-100 DLA sample combined with simultaneous observations of Ly$\alpha$ and Mg\sion{} $\lambda$ 2796 \AA{} 
provide an excellent test of the effectiveness of the Mg\sion{} selection technique for
comparison with low redshift statistics.

\section{Data}
The XQ-100 Legacy Survey observed 100 QSOs with the X-Shooter spectrograph on the Very Large Telescope, providing simultaneous wavelength coverage from $3150$ \AA{}--$25000$ \AA{} at a full width at half maximum (FWHM) resolution R$\sim$5000--9000. For more details on the observations, see \cite{Lopez16}. We emphasize that the 100 QSO targets were not pre-selected to contain DLAs, thus providing a `blind' sample of DLAs along the lines of sight. 

\cite{SanchezRamirez16} identified 41 DLAs by their Lyman series absorption in the XQ-100 spectra. However, five of these DLAs are within 5000 \kms{} of the rest-frame of the QSO. These proximate absorbers likely trace a different population of systems compared to their intervening counterparts, \citep{Ellison02,Ellison10,Berg16} and are typically ignored when computing \omegaDLA{}. We therefore restrict the DLA sample used in this Letter only to intervening DLAs. 

Table \ref{tab:EWs} contains a summary of the intervening DLAs, including the measured rest-frame EW for the Mg\sion{} 2796 \AA{} and 2803 \AA{} lines, as well as the Fe\sion{} 2600 \AA{} line (with EW \Feeqw{}). The redshift, metallicity, and \vninety{}\footnote{\vninety{} measures the velocity width corresponding to 90\%\ of the integrated optical depth using one low-ion transition \citep{Prochaska97}.} measurements are taken from \cite{SanchezRamirez16} and \cite{Berg16}. Absorption line profiles of the Mg\sion{} lines are provided in \cite{Berg16}. Additionally, we tabulated the $D$-index defined in \cite{Ellison06} and \cite{Ellison09}. For the rest of this paper, only the \ndla{} DLAs with EWs that are not blended (i.e. are not upper limits in Table \ref{tab:EWs}) are used in the analysis. 

\begin{table*}
\begin{center}
\caption{DLA properties and equivalent widths}
\label{tab:EWs}
\begin{tabular}{lccccccccc}
\hline
QSO& \zem{}& \zabs{}& logN(\HI{})& W$^{2796}_{0}$& W$^{2803}_{0}$& W$^{2600}_{0}$& [M/H] (elem)& \vninety{}& $D$-index (cut$^{a}$)\\
& & & & \AA{}& \AA{}& \AA{}& & \kms{}& \\
\hline
J0003-2603& 4.12& 3.3900& $21.40\pm0.10$& $1.393\pm0.010$& $1.147\pm0.009$& \nodata{}& $-1.93\pm0.12$ (ZnII)& 21& 5.6 (3.8)\\
J0006-6208& 4.44& 3.2030& $20.90\pm0.15$& $0.553\pm0.027$& $<0.537$& $0.383\pm0.088$& $-2.31\pm0.15$ (FeII)& 43& 5.8 (3.8)\\
J0006-6208& 4.44& 3.7750& $21.00\pm0.20$& $1.196\pm0.023$& $0.963\pm0.026$& $0.724\pm0.034$& $-0.94\pm0.20$ (ZnII)& 54& 7.0 (3.8)\\
J0034+1639& 4.29& 3.7525& $20.40\pm0.15$& $0.518\pm0.013$& $0.510\pm0.022$& $0.369\pm0.015$& $-1.88\pm0.16$ (FeII)& 32& 5.4 (3.8)\\
J0113-2803& 4.31& 3.1060& $21.20\pm0.10$& $4.377\pm0.047$& $3.620\pm0.040$& $2.326\pm0.035$& $-1.11\pm0.10$ (SiII)& 164& 8.9 (3.9)\\
J0124+0044& 3.84& 2.2610& $20.70\pm0.15$& $1.673\pm0.009$& $1.497\pm0.009$& $1.054\pm0.012$& $-0.85\pm0.15$ (SiII)& 98& 7.6 (4.2)\\
J0132+1341& 4.15& 3.9360& $20.40\pm0.15$& \nodata{}& \nodata{}& $0.235\pm0.035$& $-2.04\pm0.15$ (SiII)& 43& \nodata{} (3.8)\\
J0134+0400& 4.18& 3.6920& $20.70\pm0.10$& $0.455\pm0.008$& $0.405\pm0.007$& $0.163\pm0.015$& $-2.41\pm0.16$ (FeII)& 21& 6.0 (3.8)\\
J0134+0400& 4.18& 3.7725& $20.70\pm0.10$& $2.331\pm0.012$& $2.110\pm0.013$& $1.527\pm0.012$& $-0.91\pm0.10$ (SiII)& 98& 5.8 (3.8)\\
J0234-1806& 4.30& 3.6930& $20.40\pm0.15$& $2.422\pm0.054$& $2.101\pm0.065$& $1.063\pm0.050$& $-1.31\pm0.15$ (FeII)& 184& 5.5 (3.8)\\
J0255+0048& 4.00& 3.2555& $20.90\pm0.10$& $2.620\pm0.026$& $2.364\pm0.023$& $1.622\pm0.036$& $-1.08\pm0.10$ (SiII)& 208& 7.7 (3.8)\\
J0255+0048& 4.00& 3.9145& $21.50\pm0.10$& \nodata{}& \nodata{}& $0.457\pm0.052$& $-1.92\pm0.11$ (SII)& 21& \nodata{} (3.8)\\
J0307-4945& 4.72& 3.5910& $20.50\pm0.15$& $1.426\pm0.015$& $1.191\pm0.012$& $0.750\pm0.008$& $-1.48\pm0.15$ (FeII)& 70& 3.9 (3.8)\\
J0307-4945& 4.72& 4.4665& $20.60\pm0.10$& $1.742\pm0.020$& $1.594\pm0.025$& \nodata{}& $-1.52\pm0.10$ (SiII)& 219& 5.7 (3.8)\\
J0415-4357& 4.07& 3.8080& $20.50\pm0.20$& $<2.100$& $>1.647$& $0.755\pm0.043$& $-0.28\pm0.20$ (ZnII)& 131& \nodata{} (3.8)\\
J0424-2209& 4.33& 2.9825& $21.40\pm0.15$& $0.861\pm0.025$& $>0.824$& $0.336\pm0.024$& $-1.86\pm0.16$ (ZnII)& 32& 6.5 (3.9)\\
J0529-3552& 4.17& 3.6840& $20.40\pm0.15$& $0.375\pm0.025$& $0.087\pm0.060$& \nodata{}& $-2.38\pm0.16$ (SiII)& 21& 4.9 (3.8)\\
J0747+2739& 4.13& 3.4235& $20.90\pm0.10$& $1.601\pm0.021$& $1.431\pm0.031$& $1.635\pm0.020$& $-1.58\pm0.10$ (FeII)& 120& 7.7 (3.8)\\
J0747+2739& 4.13& 3.9010& $20.60\pm0.15$& \nodata{}& \nodata{}& $0.569\pm0.026$& $-2.03\pm0.15$ (SiII)& 153& \nodata{} (3.8)\\
J0818+0958& 3.66& 3.3060& $21.00\pm0.10$& $1.776\pm0.025$& $1.306\pm0.022$& \nodata{}& $-1.51\pm0.10$ (SiII)& 76& 3.9 (3.8)\\
J0920+0725& 3.65& 2.2380& $20.90\pm0.15$& $1.390\pm0.012$& $<1.470$& $0.864\pm0.012$& $-1.55\pm0.15$ (FeII)& 120& 7.4 (4.2)\\
J0955-0130& 4.42& 4.0245& $20.70\pm0.15$& \nodata{}& \nodata{}& $1.197\pm0.043$& $-1.54\pm0.15$ (FeII)& 336& \nodata{} (3.8)\\
J1020+0922& 3.64& 2.5920& $21.50\pm0.10$& \nodata{}& \nodata{}& \nodata{}& $-1.75\pm0.11$ (SiII)& 76& \nodata{} (4.0)\\
J1024+1819& 3.52& 2.2980& $21.30\pm0.10$& $0.839\pm0.012$& $<0.873$& $0.631\pm0.009$& $-1.45\pm0.10$ (SiII)& 54& 7.6 (4.2)\\
J1057+1910& 4.13& 3.3735& $20.30\pm0.10$& $2.456\pm0.095$& $1.956\pm0.065$& \nodata{}& $-1.21\pm0.14$ (FeII)& 153& 6.8 (3.8)\\
J1058+1245& 4.34& 3.4315& $20.60\pm0.10$& $1.673\pm0.020$& $1.330\pm0.045$& \nodata{}& $-1.85\pm0.10$ (SiII)& 127& 7.3 (3.8)\\
J1108+1209& 3.68& 3.3965& $20.70\pm0.10$& $0.325\pm0.012$& $0.274\pm0.027$& $0.235\pm0.010$& $-2.49\pm0.17$ (FeII)& 32& 4.3 (3.8)\\
J1108+1209& 3.68& 3.5460& $20.80\pm0.15$& $1.810\pm0.019$& $1.806\pm0.022$& \nodata{}& $-1.15\pm0.15$ (SII)& 70& 7.9 (3.8)\\
J1312+0841& 3.73& 2.6600& $20.50\pm0.10$& $1.237\pm0.044$& $0.860\pm0.032$& $0.548\pm0.035$& $-1.50\pm0.10$ (FeII)& 153& 5.4 (4.0)\\
J1421-0643& 3.69& 3.4490& $20.30\pm0.15$& $0.949\pm0.017$& $0.786\pm0.019$& $0.320\pm0.024$& $-1.40\pm0.16$ (FeII)& 43& 4.5 (3.8)\\
J1517+0511& 3.56& 2.6885& $21.40\pm0.10$& $0.807\pm0.021$& $0.770\pm0.023$& $0.668\pm0.007$& $-2.06\pm0.11$ (SiII)& 43& 7.1 (4.0)\\
J1552+1005& 3.72& 3.6010& $21.10\pm0.10$& $0.981\pm0.019$& $0.899\pm0.021$& $0.665\pm0.009$& $-1.75\pm0.11$ (SII)& 21& 7.4 (3.8)\\
J1633+1411& 4.37& 2.8820& $20.30\pm0.15$& $0.366\pm0.015$& $0.290\pm0.019$& \nodata{}& $-1.78\pm0.16$ (FeII)& 15& 4.8 (3.9)\\
J1723+2243& 4.53& 3.6980& $20.50\pm0.10$& $3.888\pm0.017$& $3.505\pm0.010$& $1.915\pm0.013$& $-1.03\pm0.10$ (FeII)& 374& 7.1 (3.8)\\
J2239-0552& 4.56& 4.0805& $20.60\pm0.10$& $>1.512$& \nodata{}& $0.614\pm0.010$& $-1.95\pm0.10$ (SiII)& 131& \nodata{} (3.8)\\
J2344+0342& 4.25& 3.2200& $21.30\pm0.10$& $1.524\pm0.018$& $1.189\pm0.017$& $1.056\pm0.023$& $-1.70\pm0.32$ (ZnII)& 54& 4.5 (3.8)\\
\hline
\end{tabular}
$^{a}$The cut is obtained from Table 2 in \cite{Ellison06}, and is based on the resolution at the Mg\sion{} 2796\AA{} line (assuming R=5300). An absorber with a D-index larger than the cut value would be considered a DLA candidate.
\end{center}
\end{table*}

\section{Results and Discussion}

The vast majority of DLAs identified at \zabs{}$\sim1$ have \Mgeqw{}$\geq0.6$ \AA{} (R00, R06).  For example, in a recent compilation of 369 Mg\sion{} systems (Rao et al. in prep), there are 70 Mg\sion{} absorbers with $0.3\leq$\Mgeqw{}$< 0.6$ \AA{}, but only one of these is a DLA (S. Rao private communication). As a result, many works have typically used \Mgeqw{}$\geq0.6$ \AA{} to pre-select potential DLA systems (R00, R06). Moreover, R06 used the EW of Fe\sion{} 2600 \AA{} (\Feeqw{}) to aid in identifying DLAs, and found that their DLA sample is confined to \fracMgFe{}$<2$, whereas their subDLAs span a larger range of \fracMgFe{}.

\begin{figure*}
\begin{center}
\includegraphics[width=\textwidth]{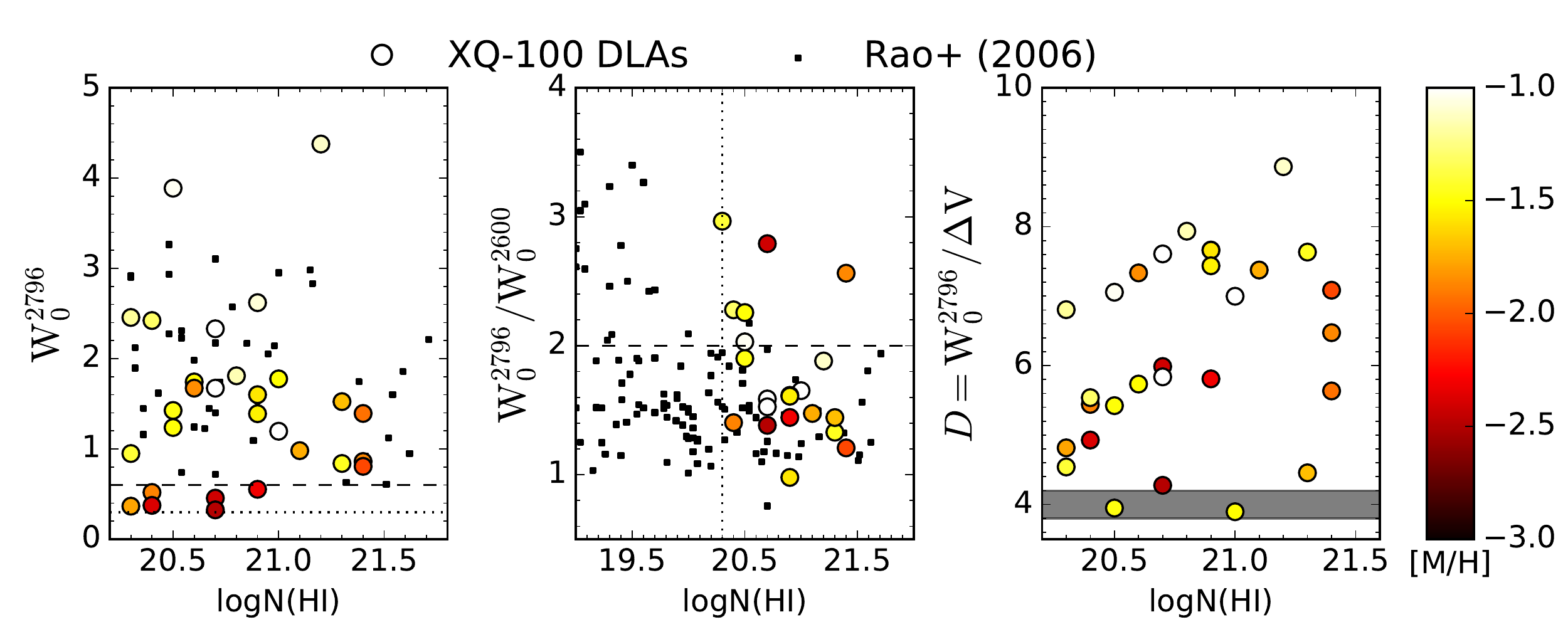}
\caption{ \emph{Left Panel}: The rest-frame \Mgeqw{} as a function of logN(\HI{}) for the XQ-100 DLAs (large circles; colours indicate metallicity) and R06 data (black squares). For reference, the \Mgeqw{} cuts (\Mgeqw{}$\geq0.3$ \AA{} and \Mgeqw{}$\geq0.6$ \AA{}) are shown as horizontal lines (dotted and dashed; respectively). Six of the \ndla{} XQ-100 DLAs show \Mgeqw{}$<0.6$ \AA{}. \emph{Middle Panel}: The ratio \fracMgFe{} as a function of logN(\HI{}) for the XQ-100 sample and R06 data. The region below the dashed line at \fracMgFe{}$\leq2.0$ characterizes the Mg\sion{}-selected DLAs in the R06 sample, while the dotted line divides subDLAs from DLAs. 30\% of the XQ-100 DLAs do not exhibit \fracMgFe{}$<$2. \emph{Right Panel}: The $D$-index (\Mgeqw{}/${\rm \Delta V}$) as a function of logN(\HI{}) for the XQ-100 DLA sample. The grey horizontal band represents the range of possible $D$-index cuts for the XQ-100 data. All XQ-100 DLAs pass their respective $D$-index cuts.}
\label{fig:MgSelection}
\end{center}
\end{figure*}

Figure \ref{fig:MgSelection} shows how the XQ-100 measurements  of \Mgeqw{} (left panel), the ratio \fracMgFe{} (middle panel), and $D$-index (right panel) vary with logN(\HI{}). Starting with the left panel of Figure \ref{fig:MgSelection}, we find that six\footnote{Two of these six DLAs with \Mgeqw{}$<0.6$ \AA{} have previously been observed in the literature (DLAs towards J1108+1209 and J0134+0400), but previous observations have not covered the Mg\sion{} absorption. We also note that one of the excluded proximate DLAs (J0034+1639 at \zabs{}$=4.25$) does not satisfy the \Mgeqw{}$\geq0.6$ \AA{} cut. This DLA is metal-poor ([M/H]$=-2.40$) and has an equivalent width \Mgeqw{}$=0.344\pm 0.013$ \AA{}.} (20\% of the sample) of the XQ-100 DLAs with measured \Mgeqw{}  have \Mgeqw{}$<0.6$ \AA{} (dashed line). However, \emph{all DLAs pass the \Mgeqw{}$\geq0.3$ \AA{} cut}.  The middle panel of Figure \ref{fig:MgSelection} shows the \fracMgFe{} ratio for the XQ-100 and R06 samples, and demonstrates that 30\% of the XQ-100 DLAs with Fe\sion{} 2600 \AA{} measurements  have \fracMgFe{}$>2.0$ (DLAs above dashed line).  Only one DLA (J0034+1639, \zabs{}$=3.69$) does not satisfy both \Mgeqw{}$\geq0.6$ \AA{} and  \fracMgFe{}$<2$ restrictions. Lastly, the right panel of Figure \ref{fig:MgSelection} shows the $D$-index \citep[i.e. \Mgeqw{} normalized by the velocity width of the line ${\rm \Delta V}$; from][]{Ellison06} for the XQ-100 DLAs as a function of logN(\HI{}). The minimum $D$-index cuts required for absorbers to be DLA candidates are derived from \citet[their Table 2]{Ellison06}, and are tabulated in Table \ref{tab:EWs}. The $D$-index cuts are based on the resolution of the X-Shooter spectrum at each Mg\sion{} line (assuming a FWHM resolution of R$=5300$). We note that the $D$-index of the XQ-100 DLAs does recover all the DLAs (i.e. points are above the grey band in the right panel of Figure \ref{fig:MgSelection}).

\emph{Are the properties of \Mgeqw{}$<0.6$ \AA{} DLAs different to the higher EW systems?}\quad The colour bar in Figure \ref{fig:MgSelection} indicates the metallicity of each DLA in the XQ-100 sample \citep{Berg16}.  Interestingly, the DLAs whose \Mgeqw{}$<0.6$ \AA{} all have low metallicities, below [M/H]$<-1.7$. The mean metallicity  of the XQ-100 DLAs subsample with \Mgeqw{}$\geq0.6$ \AA{} is [M/H]$=-1.42\pm0.03$\footnote{All errors for mean quantities are derived using a bootstrap technique with one million iterations. The errors on individual measurements were assumed to be Gaussian.}, whereas the entire sample has a mean metallicity of [M/H]$=-1.60\pm0.02$. However, the \HI{}-weighted metallicity that is generally used to trace the evolution of DLA metallicity with cosmic time \citep[e.g.][]{Pettini99,Rafelski12} is negligibly affected, increasing from [M/H]$=-1.47\pm0.03$ for the full sample to [M/H]$=-1.43\pm0.03$ when \Mgeqw{}$<0.6$ \AA{} absorbers are excluded.

\begin{figure*}
\begin{center}
\includegraphics[width=\textwidth]{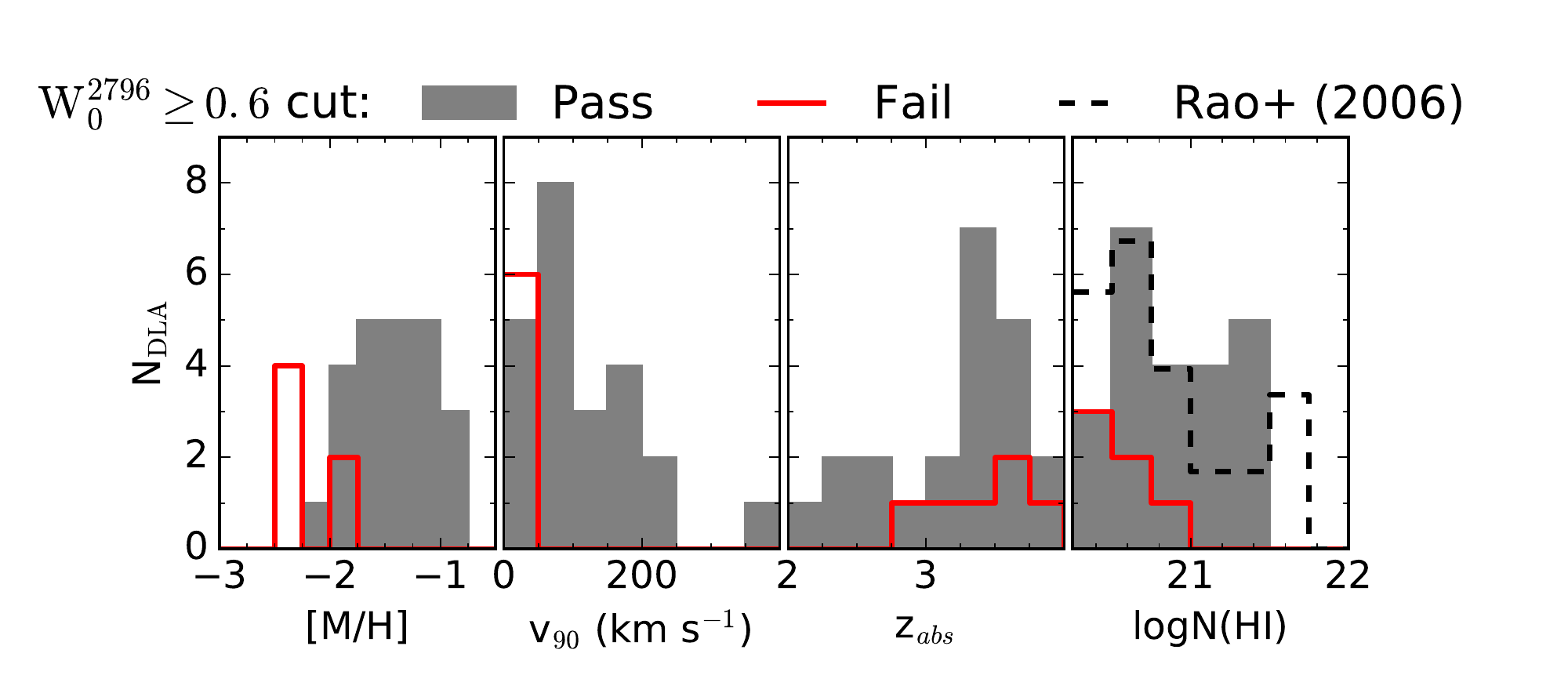}
\caption{XQ-100 DLA distribution of metallicity ([M/H]), \vninety{}, \zabs{}, and logN(\HI{}) (from left to right). Different histograms are shown for DLAs that pass (shaded grey region) or fail (red line) the equivalent width cut \Mgeqw{}$\geq0.6$ \AA{}. The simple \Mgeqw{}$\geq0.6$ \AA{} cut clearly misses low metallicity systems, with small logN(\HI{}) and \vninety{}. For reference, the \HI{} distribution of the R06 DLAs is shown as the dashed line. For visual purposes the R06 distribution is scaled down by a factor of $\sim1.8$ to match the number of DLAs in the shaded region.}
\label{fig:dists}
\end{center}
\end{figure*}

In addition to the metallicity, we checked for other DLA properties dependent on \Mgeqw{}$\geq0.6$  \AA{} selection. Figure \ref{fig:dists} shows the distributions of [M/H], \vninety{} \citep[which has been suggested as a proxy for mass, e.g.][]{Prochaska97,Haehnelt98}, \zabs{}, and logN(\HI{}) for the XQ-100 DLAs for a selection cut  of \Mgeqw{}$\geq0.6$ \AA{}. DLAs passing the equivalent width selection cut are shown as the shaded region, whilst DLAs that fail the selection cut are shown as the red line. DLAs with \Mgeqw{}$<0.6$ \AA{} tend to show low metallicities, low logN(\HI{}), and  low \vninety{} widths with respect to DLAs with \Mgeqw{}$\geq0.6$ \AA{}. These properties are consistent with a `mass-metallicity' relationship seen in DLAs \citep{Ledoux06,Jorgenson10,Moller13,Neeleman13, Christensen14}, where narrower metal lines are typically found in lower metallicity systems. The dependence of \Mgeqw{} on velocity width has been previously identified in other works \citep[][R06]{Nestor03,Ellison06,Murphy07}. The $D$-index defined in \cite{Ellison06} potentially corrects for this bias towards low \vninety{}, as the EW is normalized by the velocity width of the line. As demonstrated by the right panel of Figure \ref{fig:MgSelection}, the $D$-index provides a more complete DLA sample relative to a fixed Mg\sion{} EW cut by including those absorbers with low metallicity and \vninety{}.

\emph{What are the implications for the cosmological context of DLAs at $2\leq$\zabs{}$\leq4$?}\quad The typical approach to calculating \omegaDLA{} at high redshifts is to sum the total N(\HI{}) observed over the total redshift path (X) for all QSOs observed, i.e.

\begin{equation}
\label{eq:sum}
\Omega_{\rm HI} = \frac{H_{0} \mu m_{H}}{c \rho_{crit}} \frac{\sum N(\HI{})}{\sum \rm{X}}.
\end{equation} 

If DLAs are missed from a Mg\sion{}-selected sample, the computed \omegaDLA{} from Eq. \ref{eq:sum} would be underestimated, as the sum of N(\HI{}) would exclude the low Mg\sion{} EW DLAs while $\sum \rm{X}$ remains unaffected. For the entire XQ-100 DLA sample, \omegaDLA{} would be \emph{underestimated} by $\sim5$\% if \Mgeqw{}$<0.6$ \AA{} DLAs were excluded.

However, R06 used a different approach to compute \omegaDLA{}, that combines the number density of DLAs ($n_{\rm DLA}$; observed along Mg\sion{} absorber sightlines in R06) and the average N(\HI{}) of DLAs ($\langle {\rm N(\HI{})} \rangle$),
\begin{equation}
\label{eq:Mg}
\Omega_{\rm HI}(z) = \frac{H_{0} \mu m_{H}}{c \rho_{crit}} \frac{E(z)}{(1+z)^{2}} n_{\rm DLA}(z) \langle N(\HI{}) \rangle. 
\end{equation}

With the R06 formalism, the calculation of \omegaDLA{} depends on two measured variables: the frequency of absorbers  and their mean N(\HI{}). As shown in Fig. \ref{fig:dists}, in the XQ-100 sample, DLAs with \Mgeqw{}$<0.6$ \AA{} tend to have lower N(\HI{}) than higher EW absorbers.  If the low EW systems were not included in the XQ-100 sample statistics then  $\langle$N(\HI{})$\rangle$ and thus \omegaDLA{} would be overestimated.  However, this effect in our sample is minimal: the mean log($\langle$ N(\HI{}){}$\rangle$) of the XQ-100 sample increases minimally  from  $20.98\pm0.03$ to $21.01\pm0.03$  ($\sim7$\%) upon exclusion of the DLAs with \Mgeqw{}$<0.6$ \AA. Therefore, for a constant $n_{\rm DLA}$, \omegaDLA{} would be \emph{overestimated} by $\sim7$\% when low EW systems are excluded.

\emph{Comparison with the properties of Mg\sion{} in DLAs at \zabs{}$\leq1.5$.}\quad In the latest compilation of 369 \zabs{}$\sim1$ Mg\sion{} absorbers (Rao et al.~in prep) only 1 out of 70 (1.4$_{-1.2}^{+3.3}$\%\footnote{The subscript and superscript represent the Poisson 1$\sigma$ confidence limits derived from Tables 1 and 2 in \cite{Gehrels86}.}) systems with $0.3\leq$\Mgeqw{}$<0.6$ \AA{} is confirmed to be a DLA (S. Rao, private communication).  In contrast, $\sim7_{-2}^{+4}$\% of  $0.3\leq$\Mgeqw{}$<0.6$ \AA{} systems are DLAs\footnote{We note that the DLA incidence rate for \Mgeqw{}$>0.6$ \AA{} absorbers in the XQ-100 sample is $\sim14_{-3}^{+4}$\% (compared to  the $\sim22$\% seen at \zabs{}$\sim1$; R06).} (Lopez et al.~in prep). The DLA incidence for the low \Mgeqw{} regime is a factor of $\sim 5$ higher at \zabs{}$\sim 3$ than at \zabs{}$\sim 1$.  These incidence rates indicate a potential evolution in the Mg\sion{} properties of DLAs as a function of redshift.  Whereas DLAs at low \zabs{} are almost uniquely associated with \Mgeqw{} $\geq0.6$ \AA{}, at high redshift a significant fraction of DLAs (20\% in our sample) can have lower values. The known relationship between \Mgeqw{} and velocity spread \citep[e.g.][]{Ellison06}, as well as the relatively low values of measured \vninety{} and low metallicity of the \Mgeqw{}$<0.6$ \AA\ DLAs in the XQ-100 sample indicate that low \Mgeqw{} absorbers may preferentially be probing low mass galaxies, which are less prevalent at low redshift. This is consistent with the lack of low metallicity DLAs at low redshift \citep{Rafelski14, Berg15II}.  What is currently unknown is whether, despite their rarity, low Mg\sion{} EW  DLAs at low \zabs{} show the same distribution of properties (metallicities, \vninety{}, N(\HI{})) as high \zabs{} DLAs of the same \Mgeqw{}. Based on our high \zabs{} results, we caution that DLAs that have been selected based on a high Mg\sion{} EW cut have the potential to be biased against low metallicity systems.  Indeed, \cite{Kulkarni07} have argued that Mg\sion{} pre-selection could select against DLAs with [M/H]$<-2.5$.  Nonetheless, the low frequency of such systems at low redshifts (and their tendency towards lower N(\HI{})) means that \omegaDLA{} and the \HI{}-weighted metallicity is unlikely to be significantly affected.

\section{Summary}
Using the unbiased sample of \ndla{} DLAs from the XQ-100 sample ($2<$\zabs{}$<4$), we have investigated the Mg\sion{} properties of $2\leq$\zabs{}$\leq 4$ absorbers. In summary:

\begin{enumerate}

\item The XQ-100 DLAs span a larger range of \Mgeqw{} and \fracMgFe{} than previously seen in low-redshift samples.  20\% of the XQ-100 DLAs have \Mgeqw{}$<0.6$ \AA{}. We note that both the $D$-index presented in \cite{Ellison06} and the \Mgeqw{}$\geq0.3$ \AA{} cut (R00) identify all the DLAs in the XQ-100 sample. 

\item Using \fracMgFe{}$<2.0$ only selected 70\% of the XQ-100 DLAs, and would not aid in pre-selecting DLAs from Mg\sion{} absorbers at high redshifts. 

\item The XQ-100 DLAs with \Mgeqw{}$<0.6$ \AA{} tend to have lower metallicities, low \vninety{} and lower N(\HI{}) compared to DLAs with high Mg\sion{} EW, suggesting that DLAs at \zabs{}$\gtrsim2.0$ with \Mgeqw{}$<0.6$ \AA{} may preferentially select lower mass galaxies.  

\item The \HI{}-weighted metallicity of our complete XQ-100 DLA sample is [M/H]$=-1.47\pm0.03$, compared with [M/H]$=-1.43\pm0.03$ for only DLAs with \Mgeqw{}$\geq0.6$ \AA{}.

\item The mean N(\HI{}), and hence \omegaDLA{}, of DLAs may be over-estimated solely using \Mgeqw{}$\geq0.6$ \AA{} systems. This is due to both a bias against low N(\HI{}) absorbers and a possible over-representation of high N(\HI{}) absorbers for the \Mgeqw{}$\geq0.6$ \AA{} DLAs. If the cosmic \HI{} gas density is computed based on the summation of N(\HI{}) in the DLA sample (Eq. \ref{eq:sum}), the exclusion of \Mgeqw{}$<0.6$ \AA{} absorbers leads to a reduction in \omegaDLA{} by 5\%.  However, since DLAs associated with \Mgeqw{}$<0.6$ \AA{} absorbers tend to have lower \HI{} column densities, using the mean N(\HI{}) to compute the cosmic \HI{} gas density (e.g. Eq. \ref{eq:Mg}) results in a 7\% increase in \omegaDLA{} if \Mgeqw{}$<0.6$ \AA{} absorbers are excluded, relative to the full DLA sample.

\item There is a factor of $\sim5$ more DLAs with \Mgeqw{}$<0.6$ \AA{} at high redshifts (\zabs{}$\sim 3$) compared to lower redshifts (\zabs{}$\sim 1$), suggestive of an evolution in the Mg\sion{} properties of DLAs as a function of redshift. This evolution would be consistent with the deficit of low-metallicity systems  observed at low redshifts.

\end{enumerate}

\section*{Acknowledgements}
We are very grateful to Sandhya Rao for her useful comments on an earlier edition of this manuscript, and sharing her results of a new Mg\sion{} survey prior to publication. We also thank the referee (Michael Murphy) for his comments for improving the manuscript. SLE acknowledges the receipt of an NSERC Discovery Grant which supported this research. JXP is supported by NSF grant AST-1109447. SL has been supported by FONDECYT grant number 1140838 and partially by PFB-06 CATA. KDD is supported by an NSF AAPF fellowship awarded under
NSF grant AST-1302093.

\bibliography{bibref}
\end{document}